# Monoclinic (*Mc*) phase and electric field induced phase transformation in BaTiO$_3$

Ajay Kumar Kalyani[1], Dipak Khatua[1], B. Loukya[2], Ranjan Datta[2], Andy N.Fitch[3], Anatoliy Senyshyn[4], and Rajeev Ranjan[1*]

[1]Department of Materials Engineering, Indian Institute of Science, Bangalore-560012, India

[2]International Centre for Materials Science, Jawaharlal Nehru centre for advanced scientific research, Bangalore- 560064, India

[3]European Synchrotron Radiation Facility, BP 220 38043 Grenoble, Cedex, France

[4]Forschungsneutronenquelle Heinz Maier-Leibnitz (FRM II). Technische Universität München, Lichtenbergestrasse 1, D-85747 Garching b. München, Germany

## Abstract

For decades it has been a well-known fact that among the few ferroelectric compounds in the perovskite family namely BaTiO$_3$, KNbO$_3$, PbTiO$_3$ Na$_{1/2}$Bi$_{1/2}$TiO$_3$ the dielectric and piezoelectric properties of BaTiO$_3$ is considerably higher than the others in polycrystalline form at room temperature. Further, similar to ferroelectric alloys exhibiting morphotropic phase boundary, single crystals of BaTiO$_3$ exhibits anomalously large piezoelectric response when poled away from the direction of spontaneous polarization at room temperature. These anomalous features in BaTiO$_3$ remained unexplained so far from the structural stand point. In this work we have used high resolution synchrotron X-ray powder diffraction, atomic resolution aberration corrected transmission electron microscopy, in conjunction with a novel powder poling technique, to reveal that (i) the equilibrium state of BaTiO$_3$ is characterized by coexistence of a subtle monoclinic *(Mc)* phase and tetragonal phase, and (ii) strong electric field induces an orthorhombic phase at 300 K. These results suggest that BaTiO$_3$ at room temperature is within an instability regime, and that this instability is therefore the fundamental factor responsible for the anomalous dielectric and piezoelectric properties of BaTiO$_3$ as compared to the other homologous ferroelectric perovskite compounds. The results demonstrate that pure BaTiO$_3$ at room temperature more akin to lead-based ferroelectric alloys close to the morphotropic phase boundary where polarization rotation and field induced ferroelectric-ferroelectric phase transformations play fundamental role in influencing the piezoelectric behavior.

I. Introduction

The classical ferroelectric BaTiO$_3$ (BT) and its derivatives find extensive applications in electronic industries as multilayer capacitors and memory devices. With increasing emphasis on phasing out of toxic materials in industrial products, there is great surge in research activities on lead-free materials. BaTiO$_3$ based ceramics have recently been shown to exhibit piezoelectric properties comparable to that of the commercial piezoelectric lead-zirconate-titanate (PZT) [1] as well as giant electrocaloric response. [2] Apart from its great technological significance, the complexity associated with its structure and phase transition behaviour continues to puzzle the academic community even more than six decades after its discovery. BT exhibits three transitions with temperature at ambient pressure: cubic (Pm3m)-tetragonal (P4mm) at 130 ºC, tetragonal (P4mm) – orthorhombic (Amm2) at ~ 0 ºC and orthorhombic (Amm2) – rhombohedral (R3m) at -90 ºC with spontaneous polarization pointing along [001], [101] and [111] pseudocubic directions in the three low temperature phases, respectively. Similar to the relaxor based high performance piezoelectric systems PMN-PT and PZN-PT, single crystals of BaTiO$_3$ are known to exhibit very high piezoelectric coefficients when electrically poled along non-polar directions. [1,3] However, unlike the lead-based piezoelectrics which are essentially alloys and exhibit ferroelectric-ferroelectric instability around the morphotropic phase boundary, there has been no report of existence of intrinsic instability in BaTiO$_3$ at room temperature. Though both BaTiO$_3$ and PbTiO$_3$ are isostructural at room temperature, the dielectric and piezoelectric properties of BaTiO$_3$ is superior as compared to PbTiO$_3$. Davis et al [4] have pointed out that this behaviour is primarily due to large value of the shear piezoelectric coefficient ($d_{15}$) as compared to the longitudinal coefficient $d_{33}$ in BaTiO$_3$ as compared to PbTiO$_3$. The authors have argued that this large anisotropy arises due to proximity of the system to orthorhombic transition. In this state electric field can induce polarization rotation at relatively low energy cost. Using molecular dynamic simulations Paul et al [5] have shown that electric field induced polarization rotation in tetragonal and orthorhombic phases of BaTiO$_3$ can be facilitated by formation of intermediate monoclinic phases. In conformity with this view, a (001) field cooled BaTiO$_3$ has shown to exhibit monoclinic ($M_C$) distortion below 300 K, i.e. on approaching the stability regime of the orthorhombic phase. This field induced new phase was reported to be stable even after removal of the field [6]. Most recently, Lummen et al [7] proposed by phase-field simulations and scanning x-ray diffraction microscopy that an orthogonally domain-twinned BaTiO$_3$ would exhibit a kinetically stabilized monoclinic phase as it passes through an orthorhombic-tetragonal phase boundary. Earlier Yoshimura et al [8] observed a slightly oblique reciprocal net superposed on the (010)* section of the tetragonal reciprocal lattice in their precession photographs of $BaTiO_3$ crystal at room temperature and proposed a coherent hybrid structure comprising of tetragonal and monoclinic lattices in $BaTiO_3$ single crystals. Interestingly a similar feature observed by Matthias and von Hippel in 1948 [9] was interpreted as resulting from a twin plane. Eisenschmidt et al [10] have argued about the possibility of monoclinic distortion based on additional scattering observed between (002) and (200) tetragonal reflections when single crystal of $BaTiO_3$ was cooled towards the tetragonal-orthorhombic transition. Interestingly however a similar diffuse scattering between the 002 and 200 tetragonal peaks was interpreted by Ghosh et al [11] in their polycrystalline $BaTiO_3$ as manifestation of microstrain which increases with decreasing crystallite size. In the same work the authors reported that application of strong electric field led to development of Bragg peak at the same two-theta position which they interpreted it as signature of polymorphic transition of "unknown symmetry". The diffuse scattering at zero fields eventually turning into a Bragg peak on application of electric field seems to indicate that the diffuse peak corresponds to the precursor stage of the phase which finally acquires a long range coherence and increased latticed distortion on application of electric field. However, contrary to the observation by Cao et al [6], Ghosh et al [11] reported that the field induced new polymorph disappear after switching-off of the field. At present a coherent view is lacking which can explain all the scattered observations, sometimes seemingly contradictory, in literature. In this work, we have attempted to determine the fundamental intrinsic factor which can help not only explain the large piezoelectric and dielectric response of $BaTiO_3$ in single crystalline and polycrystalline forms, but also explain the field induced phase transition behaviour at room temperature. The problem was tackled using combined neutron powder diffraction, very high resolution synchrotron x-ray powder diffraction and aberration corrected high resolution electron microscopic imaging techniques in combination with a newly developed powder poling technique. Apart from highlighting the inadequacy of the conventional tetragonal structural model to explain specific features in the different diffraction techniques, subtle monoclinic distortion of *Mc* type could be resolved by detailed analysis of the profile shapes of higher order tetragonal Bragg reflections. An electric field driven tetragonal to orthorhombic transition was also observed at room temperature which was found to be irreversible to significant degree in free crystallites but almost completely reversible in clamped grains. Unlike in the past, where emphases have been more on establishing a correlation between extrinsic factors and the dielectric/piezoelectric properties, our results provide a structural basis for explaining the anomalous piezoelectric anisotropy in BaTiO$_3$ in terms of intrinsic polarization rotation.

## II. Experiment

BaTiO$_3$ ceramics were prepared by solid state route. High purity BaCO$_3$, TiO$_2$ were mixed in stoichiometric ratio and milled in planetary ball mill for 10 hrs. Calcination was done at 1100 C for 4hs and sintering at 1350 C for 6hrs. Neutron powder-diffraction (NPD) experiment was carried out at the SPODI diffractometer at FRM II, Germany using a wavelength of 1.548 Å [34]. High resolution synchrotron data was collected on the high resolution powder diffractometer (ID31) at European Synchrotron Radiation Facility (ESRF) France using a wavelength of 0.399959(4) Å. The powders were filled in 0.4-mm-diameter spinning borosilicate glass capillaries during the experiment. Structural refinement was carried out using Fullprof software. [12] High resolution transmission electron microscopy imaging was performed in an aberration corrected FEI TITAN 3TM 80-300 keV transmission electron microscope operating at 300 keV. A negative $C_s$ ~ -40 µm and a positive defocus of $\Delta f$ ~ + 8 nm were used to image atoms with white contrast for direct interpretation. [13–15] TEM samples from the powders were prepared first by forming a composite mass between powders and glue and then placing it between two Si supporting substrates. The sample is then thinned down mechanically to 20 µm and then Ar ion milling to perforation in order to generate electron transparent thin areas.

## III. Results

### A. *Signatures of non-tetragonal structural distortion at 300 K*

Figure 1 shows x-ray powder diffraction pattern of BaTiO$_3$ fitted with the conventional tetragonal P4mm structural model. While the Bragg positions and intensities of all the peaks could be fitted with reasonable accuracy justifying the model, a distinctly noticeable feature which could not be modeled was the additional flat intensity profile between the pair of tetragonal reflections 002 - 200 and 103-301 as shown in the insets of Fig. 1. This feature is common in the diffraction pattern of BaTiO$_3$ and is generally ignored. The additional scattering in the powder diffraction pattern stated above can be linked to the same origin which is responsible for x-ray diffuse scattering in single crystal studies. [10] Most often in powder diffraction studies such additional featureless scattering is modeled with a cubic phase which represents the disorder. For example most recently Aksel et al [16] attempted to fit the unaccounted intensities in the diffraction pattern of NBT by including a "cubic" phase in addition to the ferroelectric monoclinic phase. Earlier, Noheda et al [17] included a cubic phase to account for the additional featureless scattering in the diffraction patterns of tetragonal compositions close to the MPB of lead-zirconate-titanate. It may be remarked that this strategy is merely a matter of convenience since the true cubic perovskite phase is paraelectric in nature. Most recently, Ghost et al [11] has associated the additional scattering in $BaTiO_3$ to large microstrains in polycrystalline $BaTiO_3$. Figure 2 shows the Rietveld fitted neutron powder diffraction pattern of $BaTiO_3$ using the single phase tetragonal (P4mm) structural model. Because of the slightly inferior resolution of the neutron powder diffraction data as compared to the XRD data, the mismatch between the observed and the fitted pattern is not as pronounced in the neutron diffraction pattern at low angles. However, the benefit of being able to observe pronounced peaks at high angles in NPD offers the possibility to resolve subtle structural distortion by analyzing the profiles at scattering vectors. As shown in the inset of Fig. 2, the tetragonal Bragg peak $(214)_T$ is flanked by humps on either sides (pointed with arrows). Furthermore, there is a considerable misfit in the intensity of tetragonal $(244)_T$ and $(422)_T$ reflections. Thus, in contrast to the XRD pattern where the additional scattering was featureless, we observe additional peaks in the neutron diffraction pattern suggesting a possibility of another phase of subtle nature coexisting with the global tetragonal phase.

The indication of the additional phase apart from tetragonal was also found by negative $C_s$ HRTEM images. Figure 3 shows the HRTEM atomic imaging of $BaTiO_3$ powder specimen. In the image there are only two types of atoms 'Ba' and 'Ti' are resolved, viewing image along [110] zone axis. The schematic representations of atoms are also shown with red symbol representing 'Ba' column and florescent green representing 'Ti' column. The image has been divided into three regions. It is known that the spontaneous polarization in tetragonal $BaTiO_3$ occur by the displacement of 'Ti' along [001] direction. It can be seen in the region 1 and 3 of figure 3; 'Ti' is displaced along [001] direction. However in the region 2, 'Ti' has not displaced exactly along [001] direction (pointed with arrow) rather it is off the [001] direction. It should also be noted that this not a domain boundary because the displacement of Ti in region 1 and 3 is neither antiparallel or perpendicular to each other. This is direct evidence that at local level $BaTiO_3$ possess another structure along with tetragonal (*P4mm*). Tsuda et al [18] using convergent beam electron diffraction (CBED) analysis have reported the presence of nano-meter size local structure with rhombohedral symmetry. The local structure in region-2 shows the possibility of rhombohedral symmetry in the lattice with displacement along <111> direction (shown with arrow). However, the possibility of still lower symmetry like monoclinic cannot be ruled out, because Cao [6] and Lummen [7] have argued the presence of monoclinic (Mc) phases in single crystals by phase field simulation and optical birefringence techniques.

### B. Evidence of Phase coexistence at 300 K

To resolve the structure of the second phase accurately, high resolution synchrotron powder diffraction experiment was carried out. Fig. 4 (a-h) shows magnified plot of tetragonal $(300)_T$, $(400)_T$, $(600)_T$, $(116)_T$, $(611)_T$, $(225)_T$, $(334)_T$ and $(530)_T$ Bragg profiles. All these profiles are expected to be singlet as per the tetragonal structure. Figure (4(a-c)) shows the $\{h00\}_T$ profiles in increasing order of 'h'. It is interesting to note that while the $(300)_T$ reflection appears singlet, an asymmetry develops on the left side in $(400)_T$ which finally manifests as a new accompanying peak along with the tetragonal $(600)_T$. In fact the $(600)_T$ profile has more three peaks. Similarly the other higher order Bragg profiles (fig 4(d-h)) exhibits more number of reflections accompanying the tetragonal Bragg peak. We would like to emphasize that to ascertain the genuinity of the subtle additional peaks highlighted above- that they are not statistical fluctuations of the intensity, data was collected on the same sample after a gap of more than two months. The features were found to be exactly reproducible convincing us that the additional peaks are genuine and not an artifact of a "noisy" data. This confirmed the conclusion arrived at using the neutron powder diffraction pattern with regard to coexistence of another subtle phase. Since the number of well-defined additional peaks is more visible in the synchrotron XRD data as compared to the NPD data, structural analysis with regard to the nature of second polymorphic phase was carried out using the synchrotron XRD data.

Based on literature we considered four plausible structural models: *P4mm+Amm2, P4mm+R3m, Pm* and *P4mm +Pm* to fit the synchrotron XRD pattern. The *P4mm+Amm2* was considered because BaTiO$_3$ undergoes tetragonal to orthorhombic (Amm2) phase transition at 5 C, and that this phase may have appeared as precursor well before the thermodynamic phase transition temperature (~5 C). The possibility of *P4mm+R3m* was considered as Tsuda et al [18] have reported local rhombohedral distortion. The two low symmetry structural models *Pm* and *P4mm +Pm* were considered in view of the recent observations [10], [7]. It is important to note that in both *P4mm+Amm2* and *P4mm+R3m* models, the best fitted patterns show some Bragg positions significantly away from the profiles as marked by arrows in Fig. 5a and 5b. The single phase monoclinic (Pm) structural model was also found to be insufficient to account for the profile shapes. As with the single phase P4mm model (not shown here), the program attempted to obtain best fit for the set of multiple peaks within a given profile by increasing the width of the Bragg peaks. Of the four plausible models, the *P4mm+Pm* model gave the best fit. This model is also the most reasonable since, unlike for the P4mm+Amm2 and P4mm+R3m models, all the calculated Bragg peak lie underneath the observed profiles. The refined structural parameters of the P4mm and the Pm phases are listed in table 1. It is important to note that the monoclinic angle is very small ($\beta$ = 90.03) - the lattice of the monoclinic phase is therefore very similar to that of the tetragonal lattice. Such small distortion is obviously not possible to detect using ordinary diffraction technique and was possible due to the extraordinary resolution and high signal to noise ratio of the synchrotron XRD data. Interestingly the phase fraction of this subtle monoclinic phase is ~50 %, and this value reasonably matches with the volume fraction predicted by phase-field simulation of $BaTiO_3$ by Lummen et al. [7] These results suggest that $BaTiO_3$ at room temperature is within an instability regime, and that the large response (dielectric and piezoelectric) is due to this instability at work at room temperature. In view of the zero field instability already present at room temperature it was therefore anticipated that electric field would be able to induce phase transformation at room temperature.

### C. *An innovative new powder poling technique*

Electric field induced structural transformation study was carried out by using an innovative powder poling technique. This new technique is ex-situ based. The essential idea is to capture the residual irreversible structural change after subjecting free ferroelectric crystallites to strong electric field. Most often strong electric field is applied to dens solid specimen (single crystal or sintered ceramic discs) as they can sustain the field without dielectric breakdown. However, crystallites (grains) in dense polycrystalline specimen are clamped by the surrounding grains. For systems undergoing structural phase transitions below the sintering temperature, the spontaneous strain resulting from the phase transformation are not completely relieved in the clamped crystallites resulting in a built-in stress in the lattice. In such a scenario, the electric field induced study on dens polycrystalline ferroelectrics need not necessarily represent the electric field response of a single crystalline material with unclamped boundary. More-over diffraction data collected in in-situ electric field diffraction studies inherently contain preferred orientation effects because of orientation of ferroelectric/ferroelastic domains along the field direction. In the absence of good texture model to account for the preferred orientation effect, such data is not good for precise structural analysis. The complexity of the situation increases manifold if the system is prone to electric field driven phase transformation as it is not possible to unambiguously determine if change in the intensity at a given two-theta position in the diffraction pattern is primarily due to alteration in the domain population of the parent phase or also has contribution from a the new peak of the field induced new phase. Though, in principle orientation averaging of the diffraction patterns should be possible by rotating the specimen in all possible directions during data collection, such experiment as of now is not feasible due to experimental limitations. The nature of the field induced phase transformation on the other hand may be captured by ex-situ technique provided the residue of the transformed phase remains after switching off of the field. As most of the ferroelectric-ferroelectric phase transitions are first order in nature, the possibility of stabilizing the field induced phase as metastable phase after removal of the field offers an interesting opportunity to understand the nature of the transformed phase by conventional powder diffraction technique using powder specimen which is intrinsically orientation-averaged. The trivial strategy could be to first pole a dense polycrystalline ferroelectric specimen by application of strong electric field and subsequently grind it to obtain powder for conventional powder diffraction experiment. Due to complete randomization of the crystallites in the powder specimen the diffraction pattern would be free of complexities associated with preferred orientation induced by field and the crucial information with regard to the nature of the field induced phase and its volume fraction can be accurately determined by Rietveld analysis. The success of this approach has recently been demonstrated in $BiScO_3$-$PbTiO_3$, [19] $Na_{1/2}Bi_{1/2}TiO_3$, [20] and $Na_{0.5}Bi_{0.5}TiO_3$-$BaTiO_3$. [21] Obviously this approach would not be successful if the field induced phase disappears after removal of the field, as has been reported by Ghosh et al [11] in their in-situ electric field diffraction study on dense polycrystalline $BaTiO_3$. One plausible reason for the disappearance of the field induced phase after removal of the field could be related to the fact that the additional strain induced by the field induced phase transformation remain inside the lattice and could not be relieved due to clamped boundaries of the crystallites, and when the field is switched off this excess elastic strain assists the backward reaction and restores the state before the field was applied. If, on the other hand, the crystallite surface is unclamped (free) a fraction of the field induced transformation strain may be relieved at the crystallite surface. In such as scenario, the reveres reaction associated with the disappearance of the field induced phase after switching off of the field would remain incomplete, thereby retaining part of the field induced phase. With this in view, we developed a new technique of "powder-poling" in which free ferroelectric crystallites are subjected to strong electric field and the residual of the field induced phase is recorded and analyzed. The powder poling process involves (i) gentle mixing annealed ferroelectric powder with polymer and densification of the polymer ceramic composite by gently wetting it by a solvent which can dissolve the polymer in the composite. The dens composite can sustain song electric field much above the coercive field of the ferroelectric specimen -in the present case a field of 50 kV/cm was applied to the composite whereas the coercive field of BaTiO3 is ~ 3 kV/cm. The powder was subsequently retrieved after dissolving the polymer in the composite. The powder thus obtained is termed as "poled powder". We may add that the first step of compacting the annealed powder of $BaTiO_3$ under uniaxial load did not result in any noticeable change in the diffraction pattern. The significant change in diffraction pattern after poling the compact is therefore due to the electric field experienced by the crystallites.

### *D. Electric field induced phase transformation*

Figure 6(a-b) shows selected Bragg profiles of $BaTiO_3$ after the powder poling operation. As discussed before, annealed $BaTiO_3$ consists of coexistence of tetragonal and monoclinic phases. At the outset, the poled pattern exhibit significant broadening of the Bragg profiles as compared to the annealed pattern. Careful examination of the pattern revealed additional humps on the sides of the main reflection. For example the $\{226\}_T$ profile shows asymmetry towards left (pointed with arrow). Similarly, additional peaks in the profiles of $\{260\}_T$, $\{222\}_T$ and $\{220\}_T$ reflection implies coexistence of two/more phases in the poled specimen. Visual comparison of the annealed and the poled powder patterns clearly reveals the irreversible phase change anticipated in the free crystallite of $BaTiO_3$ and in difference with the reversible transformation observed in clamped $BaTiO_3$ grains [11]. This result clearly demonstrates the unique significance of powder poling technique. To identify the second phase in the poled specimen, careful structural analysis was carried out with Rietveld refinements by comparing the quality of fits obtained with different structural models : *P4mm, Pm, Pm+P4mm, P4mm+Amm2, P4mm+R3m,* Fig. 7(a-e). The single phase tetragonal (P4mm) and monoclinic (Pm) structural models were unsuccessful in accounting for the additional peaks. Of the three two phase structural models *P4mm+Pm, P4mm+R3m* and *P4mm+Amm2*, the most reasonable model was found to be P4mm+Amm2. The other two-phase models predicted Bragg peaks away from the observed Bragg profiles making

them unreasonable. The predicted Bragg peaks lie underneath the observed profiles for the P4mm + Amm2 phase coexistence model. Structural analysis of the poled powder specimen refinement therefore suggests that electric field induces formation of orthorhombic phase. Refined structural parameters of the poled BaTiO$_3$ are listed in table 2. Both the phases are nearly in 1:1 ratio. The presence of these two phases has also been observed in HRTEM images taken along [110], shown in figure 8(a-d). For sake of reference, Figure 8(a) shows the simulated atomic columns of Ba, Ti and O with diffraction pattern along [110] zone axis in pseudocubic symmetry. The images in Fig. 8b-d show only the 'Ti' and 'Ba' columns. The difference of tetragonal and orthorhombic can be distinguished by the spontaneous displacement of 'Ti'. In the tetragonal phase spontaneous polarization of 'Ti' is along <001>$_c$ and in orthorhombic phase along <101>$_c$ pseudocubic directions. The HRTEM atomic images show displacement of Ti in both [001] (Fig. 8c) and [10-1] directions (Fig 8d) in different regions of the lattice confirming the presence of the tetragonal and orthorhombic phases in the poled specimen.

### IV. Discussion

The presence of monoclinic (*Cm or Pm*) phase at the morphotropic phase boundary and its role in the anomalous piezoelectric response has been widely accepted in Pb based ferroelectric systems like Pb(Zr$_{1-x}$Ti$_x$)O$_3$, Pb(Mg$_{1/3}$Nb$_{2/3}$)O$_3$-PbTiO$_3$ and Pb(Zn$_{1/3}$Nb$_{2/3}$)O$_3$-PbTiO$_3$. According to polarization rotation theory low symmetry (monoclinic or triclinic) ferroelectric phase provides a continuous pathway for rotation of the electric dipole in the unit cell leading to anomalous piezo-response. Though theoretically the possibility of monoclinic distortion and polarization rotation as plausible intrinsic mechanism has been envisaged [5,22–24] for BaTiO$_3$, it has not been given serious consideration due to lack of experimental evidence. Our results provide the structural framework to support the argument by Davis et al [4] who associated the anomalous piezoelectric behaviour of BaTiO$_3$ as against the analogous isostructural compound PbTiO$_3$ to the distinctly higher shear coefficient $d_{15}$ of BaTiO$_3$. The authors categorized BaTiO$_3$ at room temperature among the family of "rotator-ferroelectrics" including PZT, PMN-PT and PZN-PT and PbTiO$_3$ as extender ferroelectric. [4] This scenario is subject to change with temperature and pressure if the thermodynamic variables push a system towards or away from the inter-ferroelectric phase transition. Thus by heating BaTiO$_3$ from above room temperature the rotator tendency is weakened and the extender tendency strengthened. For PbTiO$_3$, on the other hand, the rotator tendency can set in at high pressures. [25] Within a generalized framework of Landau-Ginzburg-Devonshire theory Budimir et al [23] have attempted to explain these phenomena in terms of anisotropic flattening of the Gibbs free-energy profile. For BaTiO$_3$ this feature was predicted to happen close to the tetragonal-orthorhombic phase transition temperature. The fact that BaTiO$_3$ is susceptible to electric field driven phase transition at room temperature confirms that the free energy barrier separating the tetragonal-orthorhombic phase fields is considerably reduced even at room temperature allowing the precursor effects to set in. The zero field monoclinic phase is manifestation of this precursor effect. Our results also validate the predictions of molecular dynamical simulations study of Paul et al [5] according to which polarization switching becomes feasible in BaTiO$_3$ at low fields through an intermediate monoclinic phase. Interestingly, Cao et al [6] did not report about monoclinic phase in the zero field state of BaTiO$_3$ at room temperature. The authors rather reported the onset of monoclinic phase in (001) field cooled BaTiO$_3$ crystal at 10 ºC and found this phase to be stable even after removal of the field. That we could see these features at room temperature may be attributed to the better resolution of our powder diffraction data. It may be noted that the orthorhombic *Amm*2 phase is the limiting case of the monoclinic *Mc* with two of the monoclinic lattice parameters of equal magnitude. The polarization in the *Mc* phase lies in the (010) plane with [001] and [101] as the limiting directions corresponding to the tetragonal P4mm and the orthorhombic Amm2 phases, respectively. Since the monoclinic *Mc* phase already exists with pseudo-tetragonal lattice parameters in the zero field cooled state of BaTiO$_3$ at room temperature, electric field is likely to rotate the polarization in the (010) pseudo tetragonal plane towards the [101]$_T$ direction. If the polarization is close to but not exactly along [101]$_T$ in the (010)$_T$ plane, the field induced phase may as well be called as pseudo-monoclinic or pseudo-orthorhombic [26]. The field induced phase transformation is an important finding in BaTiO$_3$. In previous studies related to domain engineering in BaTiO$_3$ crystals, the role of electric field was assumed to suitably align the domains favorably so as to minimize the angle between the spontaneous polarization and the field directions. [27] Though these studies were limited to establish a correlation between the domain configuration and piezoelectric property, our results and that of Cao et al [6] suggests that the poled state of BaTiO$_3$ is a two-phase state. Hence not only the domain configuration of the tetragonal phase, the interphase boundaries are therefore expected to play significantly important role in determining the piezoelectric properties in poled crystals of BaTiO$_3$. The discovery of the presence of the instability field at room in BaTiO$_3$ also helps to understand the far superior piezoelectric coefficient of BaTiO$_3$ ceramics ($d_{33}$ ~190 pC/N) as compared that of KNbO$_3$ (d$_{33}$= 97 pC/N) and (Na$_{0.5}$Bi$_{0.5}$)TiO$_3$ (d$_{33}$ ~ 70 pC/N). As compared to BaTiO$_3$, the orthorhombic phase is highly stable in KNbO$_3$ at room temperature; the tetragonal ferroelectric state occurs at ~ 215 ºC on the high temperature side, and on cooling the rhombohedral ferroelectric state sets in at ~ -63 ºC. In this context, we may note that the XRD pattern of the powder poled KNbO$_3$ was found to be exactly identical to the unpoled specimen, Fig. 9. In the case of (Na$_{0.5}$Bi$_{0.5}$)TiO$_3$ though recent studies suggested the monoclinic symmetry at room temperature [28,29] the d$_{33}$ value of NBT ceramics (70 pC/N) is far inferior compared to BaTiO$_3$. This seemingly contradictory features were recently resolved by showing that the reported monoclinic structure does not correspond to a thermodynamic state and it converts irreversibly to a single phase rhombohedral structure after poling [20,21,30].

The results presented in this paper also have relevance with regard to understanding the peculiar grain size dependence of dielectric permittivity of BaTiO$_3$ ceramics. It is known since 1950's that that polycrystalline BaTiO$_3$ exhibits a peak in relative permittivity with decreasing grain size, reaching value as high as 6000 at grains size at ~1 micron, below which it decreases again. With regard to the maximum permittivity for grain size ~ 1 micron, the two most prominent explanations are centered on (i) internal stress [31–33] and (ii) domain mobility [11,34,35] models. The internal stress model was based on the premise that the formation of stress relieving 90 degree ferroelastic domains becomes unfeasible with decreasing grain size and, hence, internal stress builds up within the grain which then drives the system towards a tetragonal-cubic transition resulting in enhanced permittivity. Though subsequent studies have not supported this model, [34,35] the presence of internal stress in influencing the dielectric and piezoelectric properties has been emphasized by Demartin and Damjanovic to explain relatively reduced domain vibration in fine grain BaTiO$_3$ as compared to coarse grained BaTiO$_3$ [36]. Ghosh et al [11], on the other hand, have recently correlated the anomalous dielectric response at 1 micron with the enhanced mobility of non-180 degree domains by in-situ electric field dependent diffraction study. Frey and Payne [37] have reported that the orthorhombic-tetragonal transition temperature in BaTiO$_3$ increases and the tetragonal-cubic transition decreases with decreasing size. In this scenario the orthorhombic phase transition temperature becomes closer to room temperature. Since formation of the monoclinic phase in the tetragonal matrix is precursor of the orthorhombic phase, the proximity of the orthorhombic transition to room temperature temperature is likely to be an important factor for the anomalous permittivity. A similar scenario is attained in coarse grained BaTiO$_3$ by dilute chemical substitutions as in dilute Zr modified BaTiO$_3$ [38-40].

## 5. Conclusion

In this paper we have shown that the classical ferroelectric compound $BaTiO_3$ does not exhibit a single phase tetragonal structure as conventionally believed for years. Using a combination of HRTEM images, neutron diffraction and high resolution synchrotron x-ray powder diffraction studies it was possible to establish that a subtle monoclinic (Pm) phase coexists with the tetragonal (P4mm) phase at room temperature. A novel technique of powder poling was adopted to demonstrate that strong electric field induces the orthorhombic phase even at room temperature. These results prove that $BaTiO_3$ at room temperature is within an instability regime, and is structurally more akin to lead based morphotropic phase boundary systems such as PZT, PMN-PT, PZN-PT instead of $PbTiO_3$. The results provide a fundamental structural framework to explain the anomalous shear piezoelectric coefficient ($d_{15}$) exhibited by $BaTiO_3$ and validate theoretical predictions with regard to possibility of stabilizing monoclinic phase in zero field state and also field induced switching to the orthorhombic state. It offers a new perspective to understand the anomalous piezoelectric and dielectric responses in poled single crystals, polycrystalline and chemically substituted $BaTiO_3$ systems which are increasingly becoming important as lead-free piezoelectric materials.

**Acknowledgment**: RR thanks the Science and Engineering Board (SERB) of the Department of Science and Technology, Govt. of India and IISC-ISRO Space Technology Cell for financial support.

## References

[1] R. Chu, Z. Xu, G. Li, H. Zeng, H. Yu, H. Luo, and Q. Yin, Appl. Phys. Lett. **86**, 012905 (2004).

[2] N. Novak, R. Pirc, and Z. Kutnjak, Phys. Rev. B **87**, 104102 (2013).

[3] S. Wada, K. Yako, H. Kakemoto, T. Tsurumi, and T. Kiguchi, J. Appl. Phys. **98**, 014109 (2005).

[4] M. Davis, M. Budimir, D. Damjanovic, and N. Setter, J. Appl. Phys. **101**, 054112 (2007).

[5] J. Paul, T. Nishimatsu, Y. Kawazoe, and U. V. Waghmare, Phys. Rev. B **80**, 024107 (2009).

[6] H. Cao, C. P. Devreugd, W. Ge, J. Li, D. Viehland, H. Luo, and X. Zhao, Appl. Phys. Lett. **94**, 032901 (2009).

[7] T. T. A. Lummen, Y. Gu, J. Wang, S. Lei, F. Xue, A. Kumar, A. T. Barnes, E. Barnes, S. Denev, A. Belianinov, M. Holt, A. N. Morozovska, S. V. Kalinin, L.-Q. Chen, and V. Gopalan, Nat. Commun. **5**, (2014).

[8] Y. Yoshimura, A. Kojima, N. Tokunaga, K. Tozaki, and T. Koganezawa, Phys. Lett. A **353**, 250 (2006).

[9] B. Matthias and A. von Hippel, Phys. Rev. **73**, 1378 (1948).

[10] C. Eisenschmidt, H. T. Langhammer, R. Steinhausen, and G. Schmidt, Ferroelectrics **432**, 103 (2012).

[11] D. Ghosh, A. Sakata, J. Carter, P. A. Thomas, H. Han, J. C. Nino, and J. L. Jones, Adv. Funct. Mater. **24**, 885 (2014).

[12] Rodrigues-J. Carvajal, *FullPROF.A Rietveld Refinement and Pattern Matching Analysis Program* (Laboratories Leon Brillouin (CEA-CNRS), France, 2000).

[13] C. L. Jia, M. Lentzen, and K. Urban, Science **299**, 870 (2003).

[14] R. Datta, S. Kanuri, S. V. Karthik, D. Mazumdar, J. X. Ma, and A. Gupta, Appl. Phys. Lett. **97**, 071907 (2010).

[15] U. Maitra, U. Gupta, M. De, R. Datta, A. Govindaraj, and C. N. R. Rao, Angew. Chem. Int. Ed. **52**, 13057 (2013).

[16] E. Aksel, J. S. Forrester, B. Kowalski, J. L. Jones, and P. A. Thomas, Appl. Phys. Lett. **99**, 222901 (2011).

[17] B. Noheda, J. A. Gonzalo, L. E. Cross, R. Guo, S.-E. Park, D. E. Cox, and G. Shirane, Phys. Rev. B **61**, 8687 (2000).

[18] K. Tsuda, R. Sano, and M. Tanaka, Phys. Rev. B **86**, 214106 (2012).

[19] Lalitha K. V., A. N. Fitch, and R. Ranjan, Phys. Rev. B **87**, 064106 (2013).

[20] B. N. Rao, A. N. Fitch, and R. Ranjan, Phys. Rev. B **87**, 060102 (2013).

[21] R. Garg, B. N. Rao, A. Senyshyn, P. S. R. Krishna, and R. Ranjan, Phys. Rev. B **88**, 014103 (2013).

[22] H. Fu and R. E. Cohen, Nature **403**, 281 (2000).

[23] M. Budimir, D. Damjanovic, and N. Setter, Phys. Rev. B **73**, 174106 (2006).

[24] D. Vanderbilt and M. H. Cohen, Phys. Rev. B **63**, 094108 (2001).

[25] Z. Wu and R. E. Cohen, Phys. Rev. Lett. **95**, 037601 (2005).

[26] M. Davis, D. Damjanovic, and N. Setter, Phys. Rev. B **73**, 014115 (2006).

[27] S. Wada, H. Kakemoto, and T. Tsurumi, Mater. Trans. **45**, 178 (2004).

[28] E. Aksel, J. S. Forrester, J. L. Jones, P. A. Thomas, K. Page, and M. R. Suchomel, Appl. Phys. Lett. **98**, 152901 (2011).

[29] S. Gorfman, A. M. Glazer, Y. Noguchi, M. Miyayama, H. Luo, and P. A. Thomas, J. Appl. Crystallogr. **45**, 444 (2012).

[30] B. N. Rao and R. Ranjan, Phys. Rev. B **86**, 134103 (2012).

[31] H. T. Martirena and J. C. Burfoot, J. Phys. C Solid State Phys. **7**, 3182 (1974).

[32] W. R. Buessem, L. E. Cross, and A. K. Goswami, J. Am. Ceram. Soc. **49**, 36 (1966).

[33] W. R. Buessem, L. E. Cross, and A. K. Goswami, J. Am. Ceram. Soc. **49**, 33 (1966).

[34] G. Arlt, D. Hennings, and G. de With, J. Appl. Phys. **58**, 1619 (1985).

[35] G. Arlt and N. A. Pertsev, J. Appl. Phys. **70**, 2283 (1991).

[36] D. Damjanovic and M. Demartin, J. Phys. Condens. Matter **9**, 4943 (1997).

[37] M. H. Frey and D. A. Payne, Phys. Rev. B **54**, 3158 (1996).

[38] A. K. Kalyani, A. Senyshyn, and R. Ranjan, J. Appl. Phys. **114**, 014102 (2013).

[39] A. K. Kalyani and R. Ranjan, J. Phys. Condens. Matter **25**, 362203 (2013)

[40] A. K. Kalyani, K. Brajesh, A. Senyshyn and R. Ranjan, Appl. Phys. Lett. **104**, 252906 (2014)

**Figures**

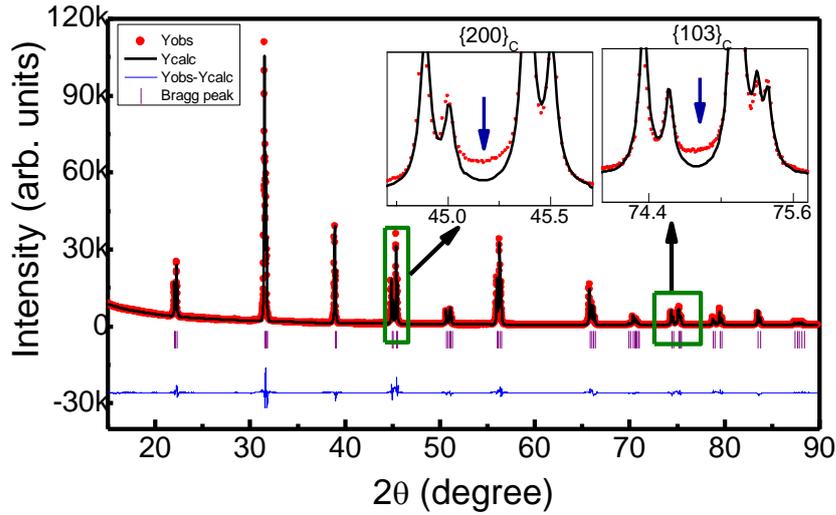

Figure 1: Rietveld fitted pattern of x-ray powder diffraction pattern of $BaTiO_3$. The insets show the misfit between the tetragonal (200) and (002), (103) and (301) reflections.

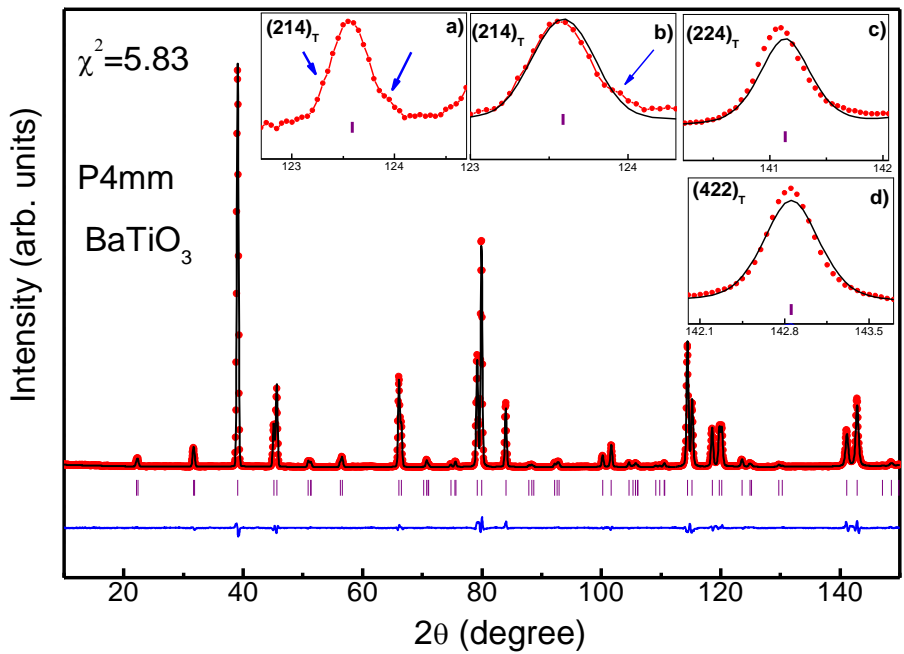

Figure 2: Rietveld fitted pattern of neutron powder diffraction pattern of $BaTiO_3$. The insets show a) pointing arrow represents the addition reflections along with the tetragonal central reflection $(214)_T$ b) misfit corresponds to the same reflection showing that the fitting has not taken into account the extra reflections. Inset c) and d) shows the misfit in the tetragonal $\{224\}_T$ and $\{422\}_T$ reflections.

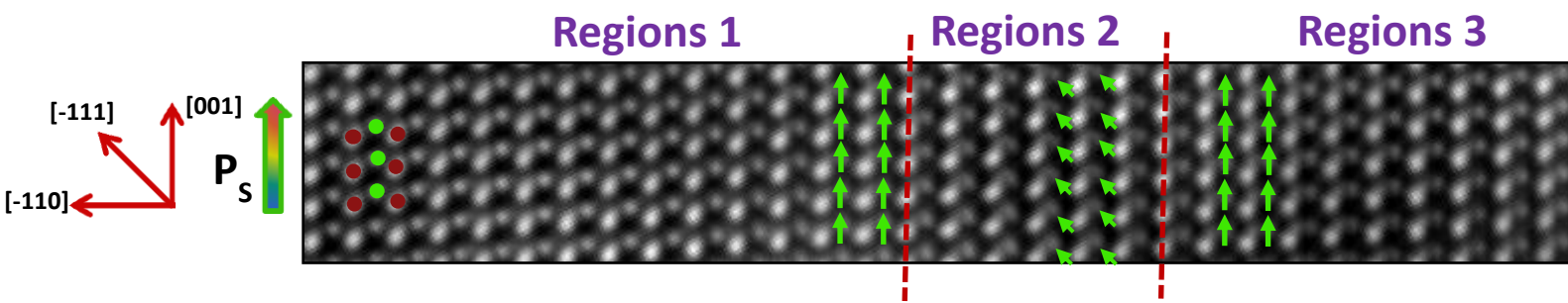

Figure 3: HRTEM image of BaTiO$_3$ with brown symbol representing 'Ba' atoms and florescent green 'Ti' atoms. The directions in the regions 1 and 3 is along [001] direction while in region 3 off [001] direction.

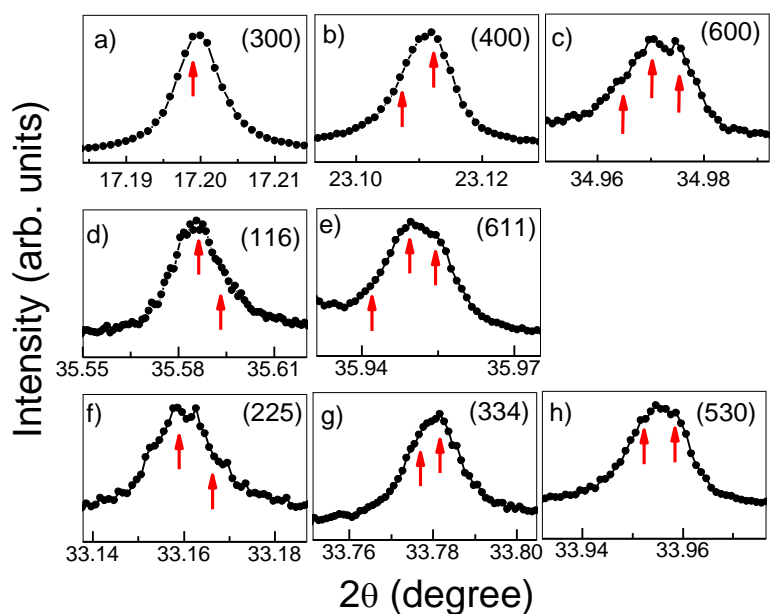

Figure 4: Synchrotron powder diffraction Bragg profiles of BaTiO$_3$. Pointing arrows represent the number of reflections in each profile.

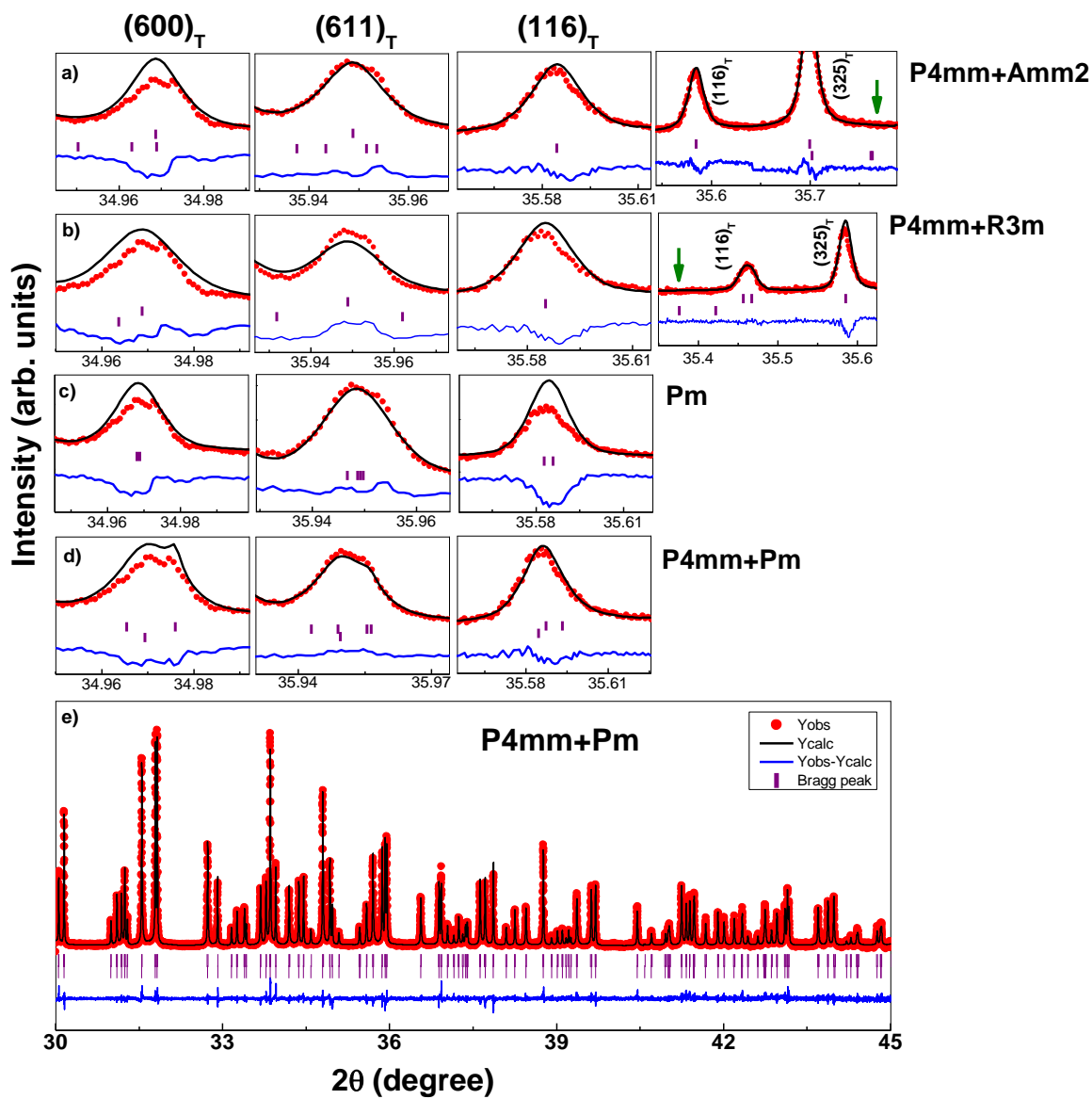

Figure 5: Rietveld fitted profiles BaTiO$_3$ refined with models; a) P4mm+Amm2, b) P4mm+R3, c) Pm, d) P4mm+Pm, e) full fitted pattern refined with P4mm+Pm space group. Arrow pointed in figure a), b) shows the peak position not accounted in the Bragg profile.

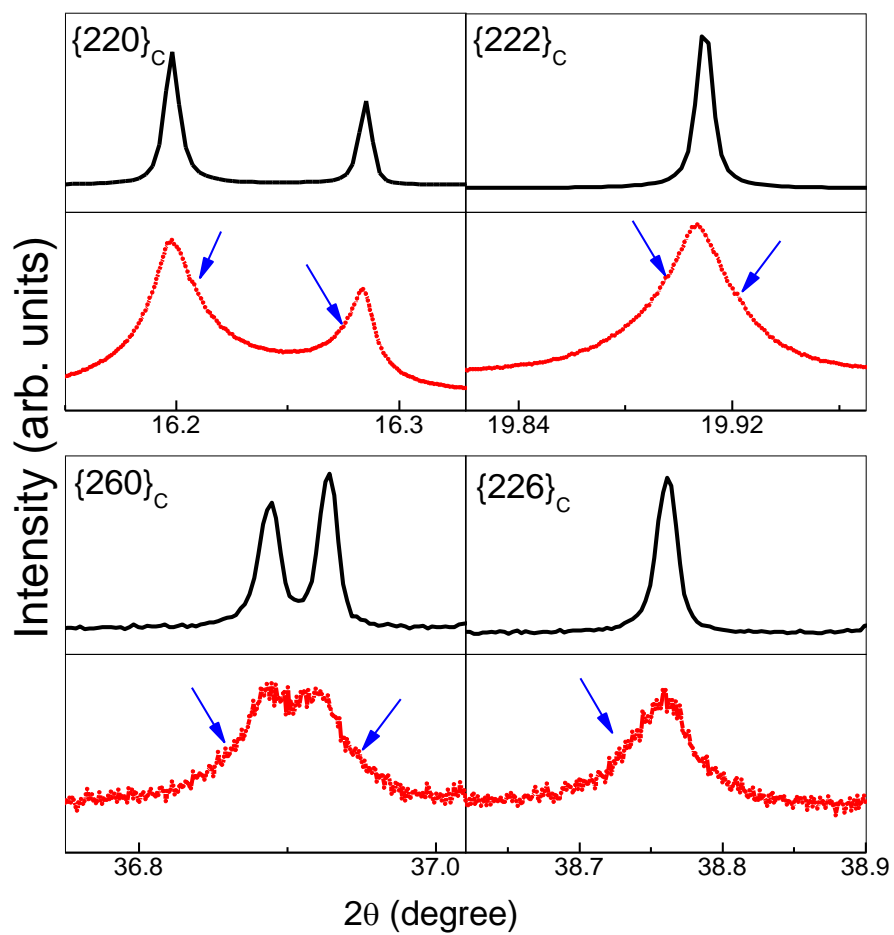

Figure 6: Synchrotron diffraction profiles of powder poled BaTiO3 in comparison with annealed powder. Pointing arrows represents the addition phase present along with the main tetragonal reflections.

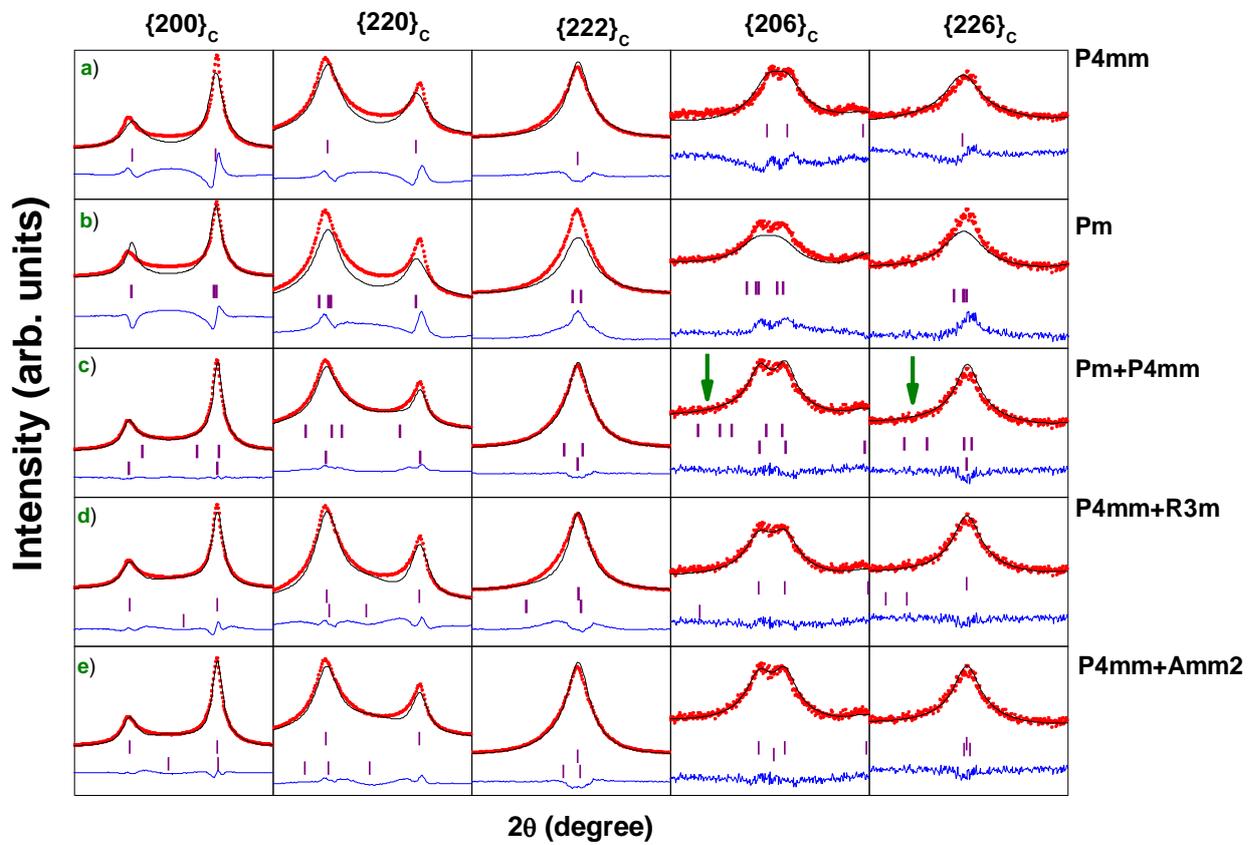

Figure 7: Rietveld fitted profiles of powder poled BaTiO3 with different structural models a) P4mm, b) Pm, c) Pm+P4mm, d) P4mm+R3m, e) P4mm+Amm2 respectively. Pointed arrow in figure c) represents the peak position accounted off the Bragg profile.

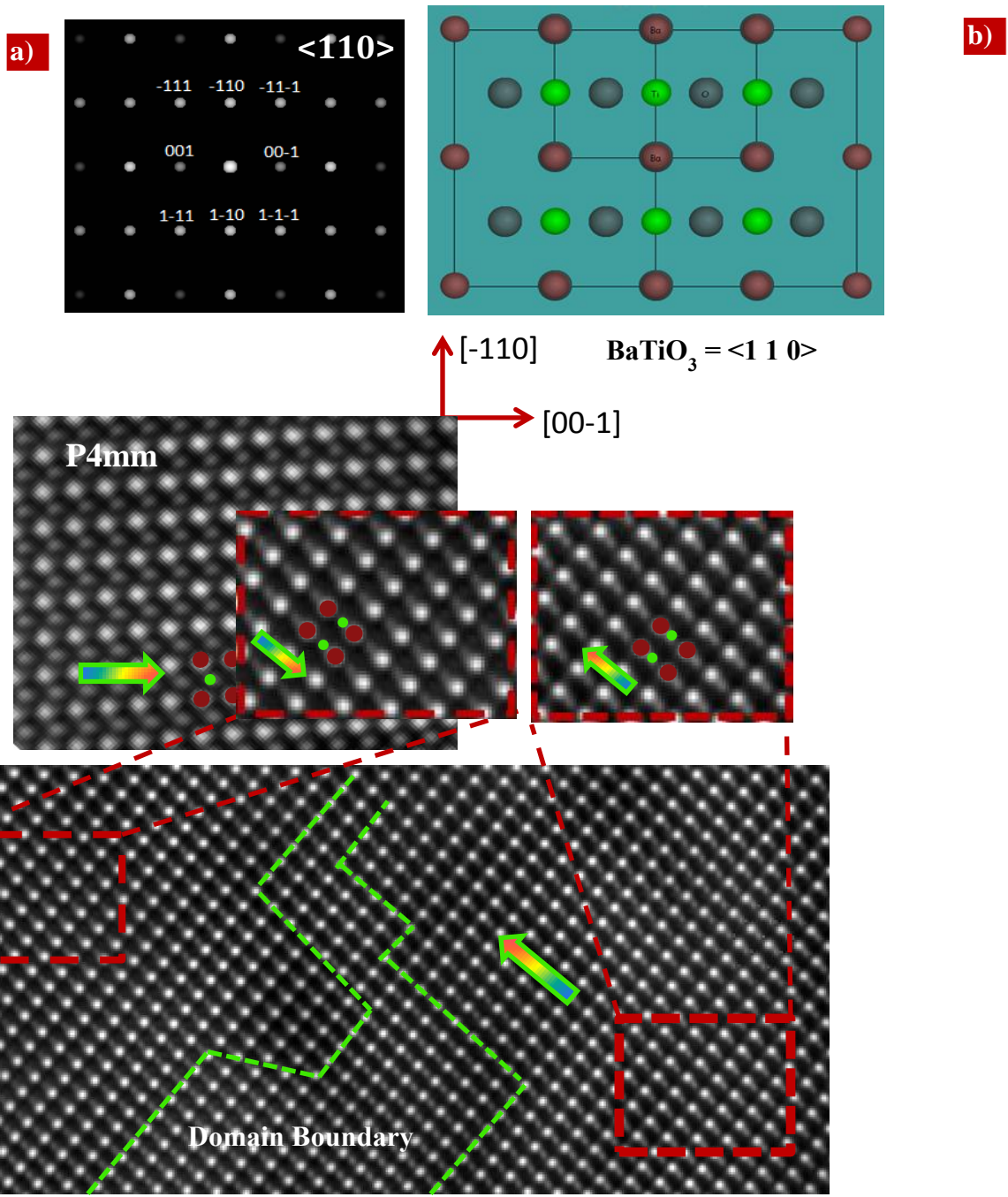

Figure 8: High resolution atomic imaging using TEM; a) simulated selected area diffraction pattern generated along [110] direction and the corresponding atomic column generated in (b). (c) Real atomic image showing the displacement of 'Ti' column along [001] direction. (d) Antiparallel displacement of 'Ti' displacement along [-110] direction with a domain boundary between them.

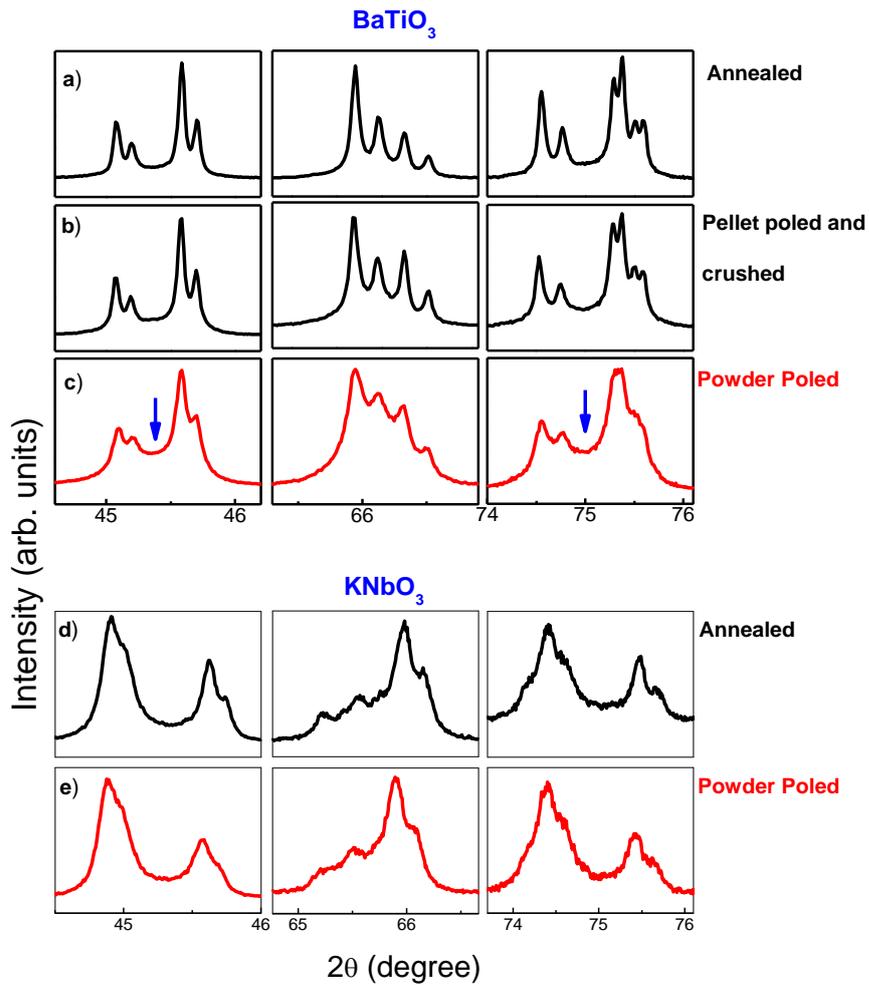

Figure 9: Shows the x-ray diffraction profiles of of BaTiO$_3$ after a) annealed, b) pellet poled and crushed and c) powder poled. Figure d) and (e) shows the x-ray diffraction profile of KNbO$_3$ after annealing and powder poled.

Table1: Refined structural parameters and agreement factors for BaTiO$_3$ using tetragonal (*P4mm*) + monoclinic (*Pm*) phase coexistence models.

| | Space group: Pm | | | | Space group: P4mm | | | |
|---|---|---|---|---|---|---|---|---|
| Atoms | x | Y | Z | B(Å$^2$) | X | y | z | B(Å$^2$) |
| Ba | 0.000 | 0.000 | 0.000 | $B_{11}$ = 0.028(1), $B_{22}$ = 0.028(1), $B_{33}$ = 0.021 (1), $B_{13}$ = -0.018(1) | 0.000 | 0.000 | 0.000 | $B_{11}$ = 0.002(3), $B_{22}$ = 0.002(2), $B_{33}$ = 0.0053 (5) |
| Ti | 0.525(1) | 0.500 | 0.484(2) | $B_{11}$ = 0.0002(1), $B_{22}$ = 0.02(1), $B_{33}$ = 0.006 (1), $B_{13}$ = 0.0032(1) | 0.500 | 0.500 | 0.522(2) | $B_{11}$ = 0.0034(3), $B_{22}$ = 0.0035(4), $B_{33}$ = 0.0031 (1) |
| O1 | 0.438(4) | 0.000 | 0.551(4) | 0.5(1) | 0.500 | 0.500 | -0.014(2) | 0.20(5) |
| O2 | 0.450(5) | 0.500 | 0.060(4) | 1.0(1) | 0.500 | 0.000 | 0.488(2) | 0.07(3) |
| O3 | -0.073(5) | 0.500 | 0.596(6) | 1.1(2) | a= 3.99359(2) Å, c= 4.03662 (2) | | | |
| a= 3.99402 (2) Å, b = 3.99286 (2) Å, c= 4.03597 (5) Å, β= 90.0291(9), v= 64.364 (1) Å$^3$, %Phase = 55(4) | | | | | v= 64.379 (2) Å$^3$, %Phase = 45(2) | | | |
| R$_p$: 17.9,  R$_{wp}$: 14.0,  R$_{exp}$: 11.58,  Chi$^2$: 1.47 | | | | | | | | |

Table 2: Refined structural parameters and agreement factors for powder poled BaTiO$_3$ using tetragonal (*P4mm*) + orthorhombic (*Amm2*) phase coexistence models.

| | Space group: P4mm | | | | Space group: Amm2 | | | |
|---|---|---|---|---|---|---|---|---|
| Atoms | x | y | Z | B(Å$^2$) | X | y | z | B(Å$^2$) |
| Ba | 0.000 | 0.000 | 0.000 | 0.104(3) | 0.000 | 0.000 | 0.000 | 0.411(6) |
| Ti | 0.500 | 0.5000 | 0.510(6) | 0.086(9) | 0.500 | 0.000 | 0.495(1) | 0.37(2) |
| O1 | 0.500 | 0.5000 | -0.013(3) | 0.25(8) | 0.000 | 0.000 | 0.478(2) | 0.32(9) |
| O2 | 0.500 | 0.000 | 0.486(2) | 0.09(3) | 0.500 | 0.255(1) | 0.231(1) | 0.02(4) |
| a= 3.99406 (1) Å, c= 4.03581 (1) Å | | | | | a= 3.99351 (3) Å, b= 5.67713 (7) Å, c= 5.68444 (7) | | | |
| v= 64.381 (1) Å$^3$, %Phase = 47.24(6) | | | | | v= 128.876 (2) Å$^3$, %Phase = 53.76(7) | | | |
| R$_p$: 6.63,   R$_{wp}$: 7.30,   R$_{exp}$: 4.19,   Chi$^2$: 3.04 | | | | | | | | |